\documentstyle [12pt] {article}

\newcommand{\cs}[3]{{{#3} \brace {#1 #2}}}

\newcommand{\h}[1]{\mathop{\lambda}\limits_{#1}\ \!\!\!}

\newcommand{\edf}{\ {\mathop{=}\limits^{\rm def}}\ }

\newcommand{\al}{\alpha}
\newcommand{\be}{\beta}
\newcommand{\m}{\mu}
\newcommand{\n}{\nu}
\newcommand{\s}{\sigma}
\newcommand{\g}{\gamma}

\newcommand{\k}{\kappa}

\begin{document}
\title{\bf NEW AND RECENT TRENDS IN MODERN COSMOLOGY}

\author{{\bf M. I. Wanas}\\
\normalsize Astronomy Department, Faculty of Science, Cairo
university, Giza, Egypt.\\ e-mail: wanas@frcu.eun.eg} \maketitle
\begin{abstract}

Non Conventional treatments in modern cosmology, in both Steady
State and Big Bang, are given. The motivation behind these
treatments is to solve some of the problems of the conventional
treatments in cosmology. For this aim, different geometric
structures and alternative field theories, used to construct world
models, are given. A brief review of Absolute parallelism (AP)
geometry and its parameterized version(PAP), as a wider geometry
than the Riemannian one, is presented. World models constructed
using alternative field theories, constructed in the AP geometry,
are discussed and compared. Some points about using topology in
the field of cosmology are commented. A new path equation,
admitted by the  PAP geometry, is used to get the effect of
spin-gravity interaction on the cosmological parameters.
\end{abstract}
\section{INTRODUCTION}

Cosmology is that branch of science which deals with the Universe
as one system. Modern Cosmology started in 1917 when A.Einstein
built his world model, in an attempt to understand the large scale
structure of the Universe (cf. [1]). This model has been built
using his theory of gravity, the {\it "General Theory of
Relativity"} (GR), which has been constructed using Riemannian
Geometry (RG). The first problem  faced by this model was that it
is not static, while it was generally believed that the Universe
is static. The model predicted that the Universe is expanding
while observations, of that time, did not support this prediction.
Einstein was enforced to modify the field equations of GR by
adding a term, called the {\it "Cosmological term"}, to these
equations, in order to stop expansion and to get a static model.
When A. Friedmann used the modified equations of GR in 1922, he
got again an expanding world model. In 1929 E. Hubble confirmed,
by studying the red-shift of distant galaxies, the prediction of
GR that the Universe is expanding.Afterwords, Einstein rejected
this term as it does not stop expansion.

After Hubble's discovery, many researches started to build models
for the Universe, in the context GR, investigating the
consequences of different assumptions about the distribution of
matter in the Universe. At that time it was generally agreed , on
observational bases, that the spatial distribution of matter in
the universe is isotropic and homogeneous. This fact represents
one of the basic assumption of modern cosmology, the {\it "
Cosmological Principle"}.

Geometers quickly entered the play ground and constructed
geometric structure that satisfy the cosmological principle [2],
[3]. They assumed that a Riemannian space whose metric is given
by, $$ ds^{2}= g_{\mu\nu} dx^{\mu} dx^{\nu}~, \eqno{(1)}$$ where $
g_{\mu\nu}$ is the metric tensor and $x^{\mu}$ is the coordinate
system used, should satisfy certain conditions (the Killing
equations), which are relations between the components of the
metric tensor and the generators of certain groups. The solutions
of the Killing equations in this case give rise to the well known
Friedmann-Robertson-Walker (FRW) metric, $$ ds^{2}= dt^{2} - \frac
{R^{2}(t)} {(1+\frac{1} {4} kr^{2})^{2}} (dr^{2}+r^{2}
d{\theta^{2}}+r^{2}{\sin^{2}\theta} d{\phi ^{2}}),~\eqno{(2)}$$
where $R(t)$ is a function of time, called the {\it "Scale
Factor"}, to be obtained from the solution of the field equations
of GR , and $ k(=+1, 0, -1)$ is the curvature constant. The
Riemannian structure given by (2) is the basis of any world model
assuming the validity of the Cosmological Principle. It is to be
considered that GR alone (i.e. without observations) cannot fix a
value for the constant k.
\subsection{Big Bang Cosmology}
Einstein's Field equations of GR can be written in the form, $$
G_{\mu\nu}= -{\kappa}T_{\mu\nu}~\eqno{(3)}$$ where $G_{\mu\nu}$ is
Einstein tensor, $T_{\mu\nu}$ is the material-energy tensor and
${\kappa}$ is Einstein's constant. Cosmology built in the context
of GR, using the FRW-space time (2) and the field equations (3),
comprises a class known as {\it "Big Bang Cosmology"}. This is
because all the solutions of (3) are singular at $t=0$, which
gives rise to the term {\it Big Bang}. The tensor $T _{\mu\nu}$ is
a phenomenological object, and not a part of the geometric
structure. Its components are chosen to satisfy the Cosmological
Principle. The simplest choice for this tensor is that of a
perfect fluid.

The Big Bang scenario has predicted that the Universe has passed
through a very hot and dense phase when it was very young. As it
expands it cools down until it reaches an estimated temperature of
$3^{0}K$ at present. Also, this scenario predicted that about $
25\% $ of the matter contents of the Universe are made of Helium
and it was formed in the first few minutes  from the Big Bang
moment, when suitable conditions for this formation were set up.
These theoretical results have been obtained by G.Gamow and his
collaborators in the mid 1940's.

\subsection{Steady State Cosmology}
Cosmological observations of 1940's show that, although the
Universe is expanding, the cosmic density of the material
distribution remains constant! Also, cosmologists using a value of
Hubble's parameter H( 130 Kms/sec/Mpc) found that the age of the
Universe is less than the age of the Earth! This represented a
puzzle for cosmologists of that time. To solve this puzzle, H.
Bondi and T. Gold in 1948 [4] suggested a different scenario for
the Universe, the {\it "Steady State Scenario"}. In the framework
of this scenario, matter  in the Universe is continuously created
as the Universe expands. This violates laws of conservation on
which GR was originally constructed. However, they rejected GR and
modified the Cosmological Principle to imply that homogeneity and
isotropy are manifested in space and time. This modification is
known as the {\it "Perfect Cosmological Principle"}. They have
built their world model without using GR or any alternative
gravity theory, as follows. In the context of the Perfect
Cosmological Principle, the curvature of space should  be constant
and since it is proportional to the quantity$ (\frac{k}{R(t)})$
they took $k=0$ to switch off the time evolution of this term.
From Hubble's diagram (the red shift- apparent magnitude
relation), they concluded that the rate of expansion of the
Universe is constant, i.e. $\frac{\dot{R}}{R}$=$H$  which has the
solution,
 $$R(t)= e^{Ht},\eqno{(4)}$$ where $\dot{R}$ is the time derivative of $R$.
This solution solved the singularity and the age problems appeared
in the Big Bang cosmology. The metric of the space (2) will now be
reduced to, $$ds^{2}= dt^{2}- e^{H t} (dr^{2}+r^{2}
d{\theta^{2}}+r^{2}{\sin ^{2} \theta}d {\phi^{2}})~.\eqno{(5)}$$

In the same year, F.Hoyle [5] modified GR field equations in an
attempt account for continuous creation of matter. He dropped
conservation by adding a term $(V_{\mu\nu})$, with non-vanishing
vectorial divergence, to Einstein's field equations(3) which he
wrote in the modified form: $$G_{\mu\nu}+ V_{\mu\nu}= -\kappa
T_{\mu\nu}~.\eqno{(6)}$$ $V_{\mu\nu}$ is a second order symmetric
tensor called the {\it "Creation Tensor"}. In 1949 Hoyle
constructed a Steady State model[6], as a solution of the modified
equations(6), assuming that the creation tensor is derived from a
vector field, called the {\it "Creation Field"}, $ V_{\mu}$ by the
relation,
$$V_{\mu\nu}\edf\frac{1}{2}(V_{\mu~;~\nu}+V_{\nu~;~\mu}).\eqno{(7)}$$
In 1960, Hoyle constructed another Steady State model assuming
that the creation tensor is derived from a scalar field which is a
function of time only. Hoyle and Narlikar constructed a steady
state model[7] as a solution of a new set of field equations
derived from an action principle, and studied its properties [8].

Each of the two rival scenarios, given above, has its own
problems. For example, the Steady State theory was ruled out in
the mid sixties since it has some problems with observations of
the cosmic microwave background radiation (CMBR) and the abundance
of light elements. On the other hand, standard Big Bang cosmology
has its own problems e.g. singularity, horizon, flatness .....
Among the history of these rival theories, scientists tried to
overcome these problems by suggesting different solutions, which
may be one or more of the following:

(1) By using other alternative theories of gravity, different from GR.

(2) By changing the basic geometry used to construct the theory.

(3) By relaxing one or more of the basic assumptions, used to
construct world models.\\ It is the aim of the present work to
give a brief review of alternative, less famous, treatments in
theoretical cosmology, which are suggested to overcome one or more
of the problems encountered by standard cosmology. In the
following sections, we are going to review briefly the
consequences of following one or more of the above suggestions. In
Section 2, a brief review of the {\it "Absolute Parallelism"}
(AP)-geometry, as an alternative to Riemannian geometry, is given.
In Section 3, two geometric AP-structures, usually used in
cosmological applications, are given. In Section 4, we review
briefly some field theories, built in AP-geometry, and give some
results of their applications in the field of cosmology. Some
comments on the possibility of using topology, in place of
geometry, are given in Section 5.
 The work is discussed in Section 6 .

\section{THE ABSOLUTE PARALLELISM GEOMETRY}
This type of geometry, the AP-geometry, was first used by
Einstein( from 1928 to 1932) in an attempt to unify gravity with
electromagnetism. As it is show, in the following brief review,
this type of geometry is more wider than the Riemannian one.
Recently, a new version of this geometry, the parameterized
Absolute Parallelism (PAP), in which both curvature and torsion
are simultaneously non-vanishing, is suggested. In the following
sub-sections, we are going to give a brief review on the AP and
PAP geometry. For more details the reader is referred to
references [9], [10], [11], [12], [13].

\subsection{The Conventional AP-Version}
The AP-space (M, $\lambda(x)$) is an n-dimensional space, each
point of which is labeled by n-variables $x^{\mu}$. Its structure
is defined completely by n- linearly independent contravariant
vectors $\h{i}^\mu (i=1,2,3,3...,n$, denotes the vector number and
$\mu =1,2,3...n$ denotes the coordinate component) defined at each
point of the manifold $ M$ and are subject to the condition, $$
 \h{i}^\m_{+|~ \n}=0,\eqno{(8)}$$
 where the stroke denotes absolute differentiation to be
 defined later. Equation (8) is the condition for the absolute
 parallelism.
 The covariant components of $\h{i}^\mu$ are defined such that,
$$ \h{i}^\mu \h{i}_\nu = \delta^{\mu}_{\nu},\eqno{(9)}$$

and
 $$\h{i}^{\nu} \h{j}_{\nu} = \delta_{ij} .\eqno{(10)}$$
 Using these vectors, the following second order symmetric
tensors are defined:$$ g^{\mu \nu} \edf \h{i}^\mu
\h{i}^{\nu},\eqno{(11)}$$ $$ g_{\mu \nu} \edf \h{i}_\mu
\h{i}_{\nu},\eqno{(12)}$$ consequently, $$ g^{\mu \alpha} g_{\nu
\alpha} = \delta^{\mu}_{\ \nu }.\eqno{(13)}$$

These second order tensors can serve as metric tensors of
Riemannian space, associated with the AP-space, when needed. This
type of geometry admits, at least, four affine connections. The
first is a non-symmetric connection given as a direct solution of
the AP-condition (8), i.e. $$ \Gamma^{\alpha}_{.~\mu \nu}\edf
\h{i}^{\alpha} \h{i}_{\mu,\nu},\eqno{(14)}$$ where the comma
denotes ordinary partial differentiation.
 The second is its dual $
\tilde{\Gamma}^{\alpha}_{.~\mu \nu}( = \Gamma^{\alpha}_{.~\nu
\mu}) $, since (14) is non-symmetric. The third one is the
symmetric part of (14),  $ \Gamma^{\alpha}_{.(\mu \nu)}$. The
fourth is Christoffel symbol defined using (11),(12) ( as a
consequence of a metricity condition). The torsion tensor is twice
the skew symmetric part of the affine connection (14), i.e. [9] $$
\Lambda^{\alpha}_{.~\mu \nu} \edf \Gamma^{\alpha}_{.~\mu \nu} -
\Gamma^{\alpha}_{.~\nu \mu}.\eqno{(15)}$$
 Another third order tensor
(contortion) is defined by the expression, $$
\gamma^{\alpha}_{.~\mu \nu} \edf \h{i}^{\alpha} \h{i}_{\mu ;
\nu},\eqno{(16)}$$ the semicolon is used for covariant
differentiation using Christoffel symbol. The two tensors are
related by, $$ \gamma^{\alpha}_{.\mu \nu}= \frac{1}{2}
(\Lambda^{\alpha}_{.\mu \nu } - \Lambda^{~ \alpha}_{\nu .\mu} -
\Lambda^{~\alpha}_{\mu .\nu}).\eqno{(17)}$$
  A basic vector could be
 obtained by contraction of the above third order tensors,$$
 C_{\mu} \edf  \Lambda^{\alpha}_{.\mu \alpha }=
\gamma^{\alpha}_{. \mu \alpha}.\eqno{(18)}$$
One of the advantages  of AP-geometry (for more details see [11]) is that for any world tensor
$T^{\alpha}_{.~\beta \gamma}$ defined in the AP-space, one can
construct a set of scalars $T_{(ijk)}$,$$
 T_{(ijk)} \edf \h{i}_{\alpha} \h{j}^{\beta} \h{k}^{\gamma}
T^{\alpha}_{.~ \beta \gamma}.\eqno{(19)}$$ If $T^{\alpha}_{.~\beta
\gamma}$ is the contortion (16) then the corresponding scalars are
those known in the literature as Ricci coefficients of rotation
[9].

The curvature tensor is defined by,$$
 B^{\al}_{.\m \n \s} {\ }  \edf {\ } \Gamma^{\al}_{.\m  \s, \n} -
\Gamma^{\al}_{.\m \n, \s} + \Gamma^{\al}_{\epsilon \n} \Gamma^
{\epsilon}_{. \m \s} - \Gamma^{\al}_{. \epsilon \s}
\Gamma^{\epsilon}_{. \m \n} \equiv 0.\eqno{(20)}$$
This tensor vanishes identically because of (8).
From the above
tensors, the following second order tensors could be defined in
Table 1 [10], [14]. \\ \\ \\ \\
\begin{center}
 Table 1: Second Order World Tensors [10], [14]      \\
\vspace{0.5cm}
\begin{tabular}{|c|c|} \hline
 & \\
Skew-Symmetric Tensors                &  Symmetric Tensors   \\
 & \\ \hline
 & \\
${\xi}_{\mu \nu} \edf \gamma^{~ ~ \alpha}_{\mu \nu .
|{\stackrel{\alpha}{+}}} $ &
\\

${\zeta}_{\mu\nu} \edf C_{\alpha}~{\gamma^{~~ \alpha}_{\mu \nu .}
} $ &
\\
 & \\ \hline
 & \\
${\eta}_{\mu \nu} \edf C_{\alpha}~{\Lambda^{\alpha}_{.\mu \nu} } $
& ${\phi}_{\mu \nu} \edf C_{\alpha}~\Delta^{\alpha}_{.\mu \nu} $
\\

${\chi}_{\mu \nu} \edf \Lambda^{\alpha}_{. \mu
\nu|{\stackrel{\alpha}{+}} }$ & ${\psi}_{\mu \nu} \edf
\Delta^{\alpha}_{. \mu \nu|{\stackrel{\alpha}{+}}} $
\\

${\varepsilon}_{\mu \nu} \edf C_{\mu | {\stackrel{\nu}{+}}} -
C_{\nu | {\stackrel{\mu}{+}}}$ & ${\theta}_{\mu \nu} \edf C_{\mu |
{\stackrel{\nu}{+}}} + C_{\nu | {\stackrel{\mu}{+}}}  $
\\

${\kappa}_{\mu \nu} \edf \gamma^{\alpha}_{. \mu
\epsilon}\gamma^{\epsilon}_{. \alpha \nu} - \gamma^{\alpha}_{. \nu
\epsilon}\gamma^{\epsilon}_{. \alpha \mu}$   & ${\varpi}_{\mu \nu}
\edf  \gamma^{\alpha}_{. \mu \epsilon}\gamma^{\epsilon}_{. \alpha
\nu} + \gamma^{\alpha}_{. \nu \epsilon}\gamma^{\epsilon}_{. \alpha
\mu}
$
\\
 & \\ \hline
 & \\
                  &  ${\omega}_{\mu \nu} \edf \gamma^{\epsilon}_{. \mu \alpha}\gamma^{\alpha}_{. \nu \epsilon}$   \\

                                      &  ${\sigma}_{\mu \nu} \edf \gamma^{\epsilon}_{. \alpha \mu} \gamma^{\alpha}_{. \epsilon \nu}$   \\

                                      &  ${\alpha}_{\mu \nu} \edf C_{\mu}C_{\nu}$   \\

                                      &  $R_{\mu \nu} \edf \frac{1}{2}(\psi_{\mu \nu} - \phi_{\mu \nu} - \theta_{\mu \nu}) + \omega_{\mu \nu}$          \\
 & \\ \hline
\end{tabular}
\end{center}
  The autoparallel
path equation can be written in the form,$$
 \frac{d^{2}x^{\mu}}{d p^{2}}+ \Gamma^{\mu}_{\alpha \beta}
\frac{d x^{\alpha} }{d p}\frac{d x^{\beta} }{d p} =
0.\eqno{(21)}$$ where $p$ is an evolution parameter.

Using the above mentioned affine connections, one can define the
following absolute derivatives [11]:$$
 A^{\m}_{+|~ \n} = A^{\m}_{~, \n} +
A^{\al}~{\Gamma}^{\m}_{.~\al \n},\eqno{(22)}$$ $$
 A^{\m}_{-| \ \n} = A^{\m}_{\ , \n} + A^{\al}
{\tilde{\Gamma}}^{\m}_{. \al \n},\eqno{(23)}$$ $$ A^{\m}_{. |~ \n}
= A^{\m}_{~, \n} + A^{\al}~{\Gamma}^{\m} _{.~(\al
\n)},\eqno{(24)}$$ $$ A^{\m}_{\  ; \n} = A^{\m}_{~, \n} + A^{\al}
\cs{\al}{\n}{\m}\,\eqno{(25)}$$ where $A^{\m }$ is any arbitrary
contravariant vector.Using these derivatives, one can define the
following curvature tensors [15], as consequence of non
commutation of these absolute derivatives,$$
\h{i}^{\stackrel{\m}{+}} _{\  | {\n \s}}  -
\h{i}^{\stackrel{\m}{+}}_ {\  | {\s \n}} \edf   \h{i}^\al B^\m_{.
\al \n \s},\eqno{(26)} $$ $$ \h{i}^{\stackrel{\m}{-}}_ {\ | {\n
\s}} - \h{i}^{\stackrel{\m}{-}}_{\  | {\s \n}}  \edf  \h{i}^\al
L^\m_{. \al \n \s},\eqno{(27)} $$ $$ \h{i}^{\stackrel{\m}{}}_ {\ |
{\n \s}}  - \h{i}^{\stackrel{\m}{}} _{\ | {\s \n}} \edf \h{i}^\al
N^\m_{. \al \n \s},\eqno{(28)} $$
 $$
\h{i}^{\mu}_{\  ; ~ \n \s}  - \h{i}^{\mu}_{\  ; ~ \s \n}  \edf
\h{i}^\al R^\m_{. \al \n \s} .\eqno{(29)}  $$ Table 2 gives a
brief comparison between the AP and the Riemannian geometries. It
shows how wide is the AP- geometry compared to the Riemannian
one.\\ \\
\begin{center}
 Table 2: Comparison Between The Riemannian Geometry and
 AP-Geometry \\
\vspace{0.5cm}
\begin{tabular}{|c|c|c|} \hline
& & \\ Object  & Riemannian geometry & AP-geometry
\\ & &
\\ \hline & & \\ Building Blocks     & $g_{\mu\nu} $
& $\h{i}^{\mu}, $
\\ & & \\ \hline & & \\ Affine Connection & $ \{^{\m}_{\al\be}\}$   &
$ \{^{\m}_{\al\be}\},
\Gamma^{\alpha}_{\mu\nu},\hat\Gamma^{\alpha}_{\mu\nu}$,
$\Gamma^{\alpha}_{(\mu\nu)}$
\\ & &
\\ \hline & &
\\ Second Order Symmetric Tensors
 & two $(g_{\mu\nu},R_{\mu\nu})$ &
 many, Table 1 \\ & & \\ \hline & &
\\ Second Order Skew Tensors &
---- & many, Table 1 \\ & & \\ \hline & & \\
 Third Order Tensor & ---- & $\gamma^{\alpha}_{.\mu\nu},\Lambda^{\alpha}_{.\mu\nu} $ \\
  & & \\ \hline & &\\
Vectors & ---- & $C_{\mu}$
\\ & & \\ \hline & & \\ Scalars & R & Many
\\ & & \\ \hline
 & & \\
Curvature& $R^{\alpha}_{.\beta\gamma\delta} \neq 0$ &
$B^{\alpha}_{.\beta\gamma\delta} \equiv  0$ \\  &  & \\ & &
$L^{\alpha}_{. \beta \gamma \delta}\neq 0 $ \\ & & \\ & &
$N^{\alpha}_{. \beta \gamma \delta} \neq 0$
\\ & &\\ & & $R^{\alpha}_{. \beta \gamma \delta} \neq 0$
\\ & & \\ \hline
\end{tabular}
\end{center}
Because of (20) many authors believe that the AP-space is a flat
one. This is not the case, since (27), (28) and (29) are
non-vanishing. For more details about this problem the reader is
referred to [12].
\subsection{The Non-Conventional PAP-Version}
There are at least two convincing physical reasons for
parameterizing the AP-geometry [11]. To clarify the first, let us
examine the structure of the curvature tensor given by (20). As
stated before, this tensor vanishes identically because of the
AP-condition (8). This tensor can be written in the form, $$
B^\al_{.\ \m \n \s} \edf R^\al_{.\ \m \n \s} + Q^\al_{.\ \m \n
\s}, \eqno{(30)} $$ where $ R^\al_{.\ \m \n \s} $ is the
Riemann-Christoffel curvature tensor, of the associated Riemannian
space, given by, $$ R^\al_{.\ \m \n \s} \edf {\cs{\m}{\s}{\al}},
\n - {\cs{\m}{\n}{\al}}, \s + \cs{\m}{\s}{\be} \cs{\be}{\n}{\al} -
\cs{\m}{\n}{\be} \cs{\be}{\s}{\al}, \eqno{(31)}$$ and $$ Q^\al_{.\
\m \s \n} \edf \g^{\stackrel{\al}{+}}_{{.\
{\stackrel{\m}{+}}{\stackrel{\n}{+}}} | \s} -
\g^{\stackrel{\al}{+}}_{.\ {{\stackrel{\m}{+}}{\stackrel{\s}{-}}}
| \n} + \g^{\be}_{. \m \s} {\ } \g^{\al}_{. \be \n} {\ } -
\g^{\be}_{. \m \n} {\ } \g^{\al}_{. \be \s}.\eqno{(32)}$$ It is
clear from (31) that $R^\al_{.\ \m \n \s}$ is made of Christoffel
symbols only, while from (32) we can see that $Q^\al_{.\ \m \n
\s}$ is made of the contortion (or the torsion via (17)) only.
Some authors believe that $R^\al_{.\ \m \n \s}$ and $Q^\al_{.\ \m
\n \s}$ are equivalent. Others consider $Q^\al_{.\ \m \n \s}$ as
giving an alternative definition of $R^\al_{.\ \m \n \s}$. Let us
examine these two tensors from a different point of view. It is
well known that Christoffel symbol is related, in applications, to
the gravitational field. So, its existence in (31) indicates that
gravity is responsible for the curvature of space-time. In our
point of view [11], the identical vanishing of the curvature
$B^\al_{.\ \m \n \s}$ may indicate that there is another physical
interaction (anti-gravity, say) which is related to the contortion
(or the torsion) and is represented by the tensor $Q^\al_{.\ \m \n
\s}$. This interaction balances the effect of gravity in such a
way that the total effect vanishes. If so, it is better to call
the tensor $Q^\al_{.\ \m \n \s}$ {\bf The Curvature Inverse of
Riemann-Christoffel Tensor}. But since gravity is dominant in our
observable Universe, which means that $ R^\al_{.\ \m \n \s} $ is
more effective than $Q^\al_{.\ \m \n \s}$, thus one has to
parameterize torsion terms in AP- expressions.

 The second reason is that the AP-geometry admits paths [16]that are
different from those of the Riemannian geometry. The new paths
contain a torsion term, together with the Christoffel symbol term.
These paths cannot be reduced to the geodesic one, unless the
torsion vanishes. It has been shown that the vanishing of the
torsion of the AP-space will reduce the space to a flat one [17].
So, what are the physical trajectories of particles that can be
represented by these paths? Clearly there are no particles that
move along the new paths. The reason is that the effect of the
Christoffel symbol term, in these equations, is of the same order
of magnitude as the effect of the torsion term. So, for these
paths to represent physical trajectories, the torsion term in the
path equations should be parameterized, in order to reduce its
effect [11].

As it is shown the two reasons for which we parameterize the
geometry are the vanishing of the curvature tensor (20) and the
problem of the physical meaning of the set of path equations
admitted by the AP-geometry. As it is clear, the common factors
between these two reasons are the affine connections. So, it is
necessary to start parameterizing these connections first.

\underline{\it Parameterized Connection}: One way to parameterize
the AP-geometry is to define a general affine connection by
linearly combining the affine connections defined in the geometry.
In doing so, we get after some manipulations [18]: $$
\nabla^\m_{.\al \be} = a_1  {\cs{\al}{\be}{\m}} + (a_2 - a_3)
\Gamma^{\m}_{. \al \be} - (a_3 + a_4)\Lambda^{\m}_{. \al \be}{\
}{\ }{\ }, \eqno{(33)} $$ where $a_1 , a_2, a_3$ and $a_4$ are
parameters. It can be easily shown that $\nabla^\mu_{.\alpha
\beta}$ transforms as an affine connection, under the group of
general coordinate transformations provided that a metricity
condition is imposed. It is clear that this parameterized
connection is non-symmetric.

\underline{\it Parameterized Absolute derivatives}: If we
characterize absolute derivatives, using the connection (33), by a
double stroke, then we can define the following derivatives: $$
A^{\m}_{+||~ \n} \edf A^{\m}_{~, \n} + A^{\al}\nabla^\m _{.\al
\n}, \eqno{(34)} $$ $$ A^{\m}_{-||~ \n} \edf A^{\m}_{~, \n} +
A^{\al}\nabla ^\m_{.\n \al}, \eqno{(35)} $$ $$ A^{\m}_{~ ||~ \n}
\edf A^{\m}_{~, \n} + A^{\al}\nabla ^\m _{.(\al \n)}, \eqno{(36)}
$$ where $A^\m$ is any arbitrary vector. The metricity, using the
parameterized connection, is given by, $$ g_{\stackrel{\m}{+}
\stackrel{\n}{+} ||\sigma} = 0, \eqno{(37)} $$ which gives rise to
the condition , $$
  a+b=1, \eqno{(38)}
$$ where  $a=a_1$, $b=a_2+a_4$, ($a_3 = -a_4$) are two parameters.
In this case the general affine connection (38) can be written in
the form: $$ \nabla^\al_{. \m \n} =  {\cs{\m}{\n}{\al}} + b
\g^\al_{. \m \n} . \eqno{(39)}$$ It is clear from this equation
that we have parameterized the contortion (or equivalently the
torsion) term in a general connection of the AP-geometry. Now we
will explore the consequences of this parameterization.

\underline{\it Parameterized Path Equation}: Using the
parameterized connection
 (39) and following the same approach followed before in [16],
 we can get the following parameterized path equation admitted by the geometry [18],
$$ {{\frac{dZ^\m}{d\tau}} + \cs{\n}{\s}{\m}\ Z^\n Z^\s = - b{\ }
\Lambda_{(\n \s)}.^\m~~  Z^\n Z^\s}, \eqno{(40)}$$ where $Z^\m$
$(\edf \frac{dx^\m}{d\tau})$ is the tangent to the path and $\tau$
is the evolution parameter along it.

\underline{\it Parameterized Curvature Tensors}: There are two
methods for defining the curvature tensor. The first is by
replacing Christoffel symbols, in the definition of
Riemannian-Christoffel curvature tensor, by any affine connection.
The second is by using the non commutation properties of the
absolute derivatives as done in Subsection 2.1. The two methods
are equivalent in RG only. Using the first method, we can define
the following curvature tensor,$$ {\hat {B}}^\al_{\ \m \n \s} \edf
\nabla^\al_{. \m \s , \n} - \nabla^\al_{. \m \n , \s} +
\nabla^\be_{. \m \s} \nabla^\al_{. \be \n} - \nabla^\be_{. \m \n}
\nabla^\al_{. \be \s}. \eqno{(41)}$$ Using the definition of
$\nabla^\be_{. \m \n} $ given by (39) then we can write, $$ {\hat
{B}}^\al_{.\ \m \n \s} = R^\al_{.\ \m \n \s} + b {\hat
{Q}}^\al_{.\ \m \n \s} , \eqno{(42)}$$ where $$ {\hat {Q}}^\al_{.\
\m \s \n} \edf \g^{\stackrel{\al}{+}}_{.\
{{\stackrel{\m}{+}}{\stackrel{\n}{+}}} | \s} -
\g^{\stackrel{\al}{+}}_{.\ {{\stackrel{\m}{+}}{\stackrel{\s}{-}}}
| \n}  + b (\g^{\be}_{. \m \s} {\ } \g^{\al}_{. \be \n} {\ } -
\g^{\be}_{. \m \n} {\ } \g^{\al}_{. \be \s} {\ }).  \eqno{(43)}$$
It is clear that the tensor ${\hat {B}}^\al_{.\ \m \n \s}$ is a
parameterized replacement of the tensor $B^\al_{.\ \m \n \s}$
given by (20). But here ${\hat {B}}^\al_{.\ \m \n \s}$ is, in
general, non-vanishing.

Using the second method, for defining curvature tensors we get the
following tensors,

$$ \h{i}^{\stackrel{\m}{+}} || {\n \s}  - \h{i}^{\stackrel{\m}{+}}
|| {\s \n}   = \h{i}^\al W^\m_{. \al \n \s}, \eqno{(44)}$$ $$
\h{i}^{\stackrel{\m}{-}} || {\n \s}  - \h{i}^{\stackrel{\m}{-}} ||
{\s \n}   = \h{i}^\al {\hat{L}}^\m_{. \al \n \s},  \eqno{(45)}$$
$$ \h{i}^{\stackrel{\m}{}} || {\n \s} - \h{i}^{\stackrel{\m}{}} ||
{\s \n}   = \h{i}^\al {\hat{N}}^\m_{. \al \n \s}. \eqno{(46)}$$

Note that every tensor with a hat is the parameterized replacement
of that without a hat. We can show that the tensors given by the
second method are more general than those obtained using the first
method. For example we can write, $$ W^\al_{. \m \n \s} = {\hat
{B}}^\al_{. \m \n \s} - b(b - 1) \g^\al_{. \m \be} \Lambda^\be_{.
\n \s}. \eqno{(47)}$$ An important results is that the
PAP-geometry is more general than both RG and AP-geometry. It
possesses curvature and torsion which are simultaneously
non-vanishing. Furthermore, from the PAP geometry, we can get RG
as a special case corresponding to $b=0$ and we can get
AP-geometry corresponding to $b=1$.
\\
\section{AP-STRUCTURES FOR COSMOLOGICAL APPLICATIONS }
In the context of gravity theories written in Riemannian geometry,
certain geometric structures are needed to construct world models,
e.g. FRW- structure (2). Similarly for field theories constructed
in AP-geometry, one needs certain AP-structures, satisfying the
cosmological principle, in order to construct world models.
Robertson [13] constructed two AP-structures for cosmological
applications. The two structures satisfy the cosmological
principle. The structure of an AP-space, of 4-dimensions, is given
by a tetrad vector field. The following two tetrad vector fields
give the complete structure of the two AP- spaces used for
cosmological applications, which can be written in spherical polar
coordinate [10], respectively as,

\underline{The 1st structure}  $$ \h{i}^{\mu} ={\left(\matrix{
  \sqrt{-1} & 0 & 0 & 0 \cr
  0 & \frac{L^+ sin\theta cos\phi}{4 R} & \frac{(L^-cos\theta cos\phi-4K^{\frac{1}{2}}r sin\phi)}{4 r R} & \frac{-(L^- sin\phi+4k^{\frac{1}{2}}r cos\theta cos\phi)}{4rR
  sin\theta}\cr
  0 & \frac{L^+ sin\theta sin\phi}{4 R}  & \frac{(L^- cos\theta sin\phi-4K^{\frac{1}{2}}r cos\phi)}{4 r R} & \frac{(L^- cos\phi-4k^{\frac{1}{2}}r cos\theta sin\phi)}{4rR sin\theta}\cr
  0 & \frac{L^+ cos\theta}{4 R} & \frac{-L^- sin\theta}{4 r R} & \frac{K^{\frac{1}{2}}}{4 R}
 }\right)}~~.\eqno{(48)}$$

\underline{The 2nd structure} $$ \h{i}^{\mu} ={\left(\matrix{
               \sqrt{-1} \frac{L^-}{L^+}    &\frac{-K^{\frac{1}{2}}r}{R}                     &0
               &0 \cr
               \sqrt{-1}(\frac{4K^{\frac{1}{2}}r sin\theta cos\phi}{L^+})    &\frac{L^- sin\theta cos\phi}{4R}
               &\frac{l^+ cos\theta cos\phi}{4rR} &\frac{-L^+ sin\phi}{4rR sin\theta}\cr
               \sqrt{-1}(\frac{4K^{\frac{1}{2}}r sin\theta sin\phi}{L^+})      &\frac{L^- sin\theta sin\phi}{4R}    &\frac{l^+ cos\theta sin\phi}{4rR}
               &\frac{L^+ cos\phi}{4rR sin\theta} \cr
               \sqrt{-1}(\frac{4K^{\frac{1}{2}}cos\theta}{L^+})    &\frac{L^- cos\theta}{4R}   &\frac{-L^- sin\theta}{4rR}   &0
 }\right)}~~.\eqno{(49)}$$ where $L^\pm = 4 \pm k r^{2},$ and
 $R(t) $ is an unknown function of t.
  It is to be considered that the Riemannian
 space associated with each one of the structures (48) and (49)
 is that given by the FRW-metric (2).

 \section{COSMOLOGICAL APPLICATIONS OF ALTERNATIVE THEORIES}
 In the present section, we are going to review briefly some alternative theories,
 different from GR, and world models resulting from the solution of their field equations .
 The general feature of these field theories is that all of them are constructed in
 spaces with absolute parallelism. Some of these theories were
 constructed to overcome one or more of the problems appeared in
 the applications of GR, especially in the cosmological case. Other
 theories are constructed to widen the domain of GR. In each of
 the following subsections, we review briefly a theory, the
 motivation for constructing its field equations and some features of  the world models
 resulting from its cosmological applications.

 \subsection{McCrea-Mikhail Treatment of Creation of Matter}
 Hoyle's modification of GR [6], to account for continuous creation of matter
 stimulated many questions. One of these questions was about the
 role of the skew part of the tensor $V_{\mu\nu}$. Another
 important questions is whether it is better to define the
 creation vector from the geometric structure used. Mikhail [10],
 McCrea and Mikhail [19] have used the AP-geometry in order to
 define this vector. They wrote
 Hoyle's modified equations (6) in AP-geometry by using the basic
 vector given by (18) to play the role of the creation vector.
 Their field equations are similar to (6). An advantage of this
 treatment is that the AP-geometry possesses sufficient structure
 to allow for the creation vector to be represented, as a
 geometric object. This represents an example for applying
 Einstein's philosophy of geometerization of physics. Applying the
 modified field equations using the geometric structures (48),
 (49), they have got a Steady State model. It is to be noted that
 the skew part of the creation tensor as defined by McCrea and Mikhail, in the above mentioned
 tetrads, vanishes identically. This is another advantage of this
 treatment.

 \subsection{A Pure Geometric Approach to the Steady State}
 Mikhail [20] constructed a Unified field theory using
 AP-geometry. This theory is pure geometric in the sense that it
 has no phenomenological objects. Its field equations can be
 written as,
  $$
\Lambda ^{\stackrel{\al}{+}}_{.\
{{\stackrel{\m}{+}}{\stackrel{\n}{+}}} | \al}= c_{\mu
 ,\nu}- c_{\nu ,\mu} = 0, \eqno{(50a)}
 $$
 $$
c_{\alpha} \Lambda^{\alpha}_{\mu\nu}= 0, \eqno{(50b)}
 $$
 $$
c_{\stackrel{\mu}{+} | \nu}+c_{\stackrel{\nu}{+}| \mu} = 0.
\eqno{(50c)}
 $$
 Applying this set of field equations to the AP-structure given by
 (48), Mikhail [21] found that the first two equations of (50) are
 satisfied identically while the last one gives,
 $$\frac{\ddot{R}}{R}-\frac{ \dot{R}^{2}}{R^{2}}=0$$ which has the
 Steady State solution,$$ R(t)=A e^{\frac{ct}{a}},\eqno{(51)}$$
 where $A$, $a$ are constants. The Riemannian space associated with
 (48) is the FRW-structure (2), which can be written in the form $$
 ds^{2}=c^{2} dt^{2}-\frac{e^{\frac{2ct}{a}}}{(1+\frac{1}{4}
 kr^{2})^{2}}(dr^{2}+r^{2}d\theta^{2}+r^{2}
 sin\theta^{2}d\phi^{2}).\eqno{(52)}$$ It is to be considered
 that, here again as in the previous subsection, the skew part
 corresponding to the creation tensor vanishes identically. This answers,
 partially, one of the questions raised above.

 One of the objections which can be raised against this treatment is
 that the field equations (50) are twenty two while the field
 variables are only sixteen (the tetrad components). Another
 objection is clear from (52) in which $k$ takes all possible
 values which may violate the perfect cosmological principle.
 However, any of the values of this constant could be inserted by
 hand, since this treatment does not fix it.

 \subsection{M\o ller's Tetrad Theory (MTT) and the Big Bang}
 In 1978 C.M\o ller [22] attempted to modify GR in order to remove the
 inevitable singularities appearing in the solutions of its field
 equations. He wrote a new gravity theory, in the AP-geometry,
 whose field equations are derivable from an action principle.
 The Lagrangian function suggested by M\o ller is in the form:
 $$\pounds = \pounds_{g}+\pounds_{m} $$  where $$
 \pounds_{g}\edf\sqrt{-g}(\alpha_{1}c_{\mu}c^{\mu}+\alpha_{2}\gamma_{\mu\nu\sigma}
 \gamma^{\mu\nu\sigma}+\alpha_{3} \gamma_{\mu\nu\sigma}
 \gamma^{\sigma\nu\mu})\eqno{(53)}$$
 $$\pounds_{m}\edf\k\sqrt{-g}T^{\mu\nu}g_{\mu\nu}\eqno{(54)}$$
 where $\alpha_{1}$, $\alpha_{2}$ and $\alpha_{3}$ are parameters
 to be fixed later, and $T^{\mu\nu}$ is a phenomenological
 material-energy tensor. M\o ller imposed the condition that his
 suggested theory must have a correct Newtonian limit. Using this
 condition, he was able to reduce the three parameters to only one
 parameter, $\chi$, where
 $$\alpha_{1}=-1~~,~~ \alpha_{2}=\chi~~,~~ \alpha_{3}=1-2\chi ~.$$.
 His field equations can be written in the form,$$
 G_{\mu\nu}+H_{\mu\nu}=-\kappa T_{\mu\nu}\eqno{(55)}$$
 $$f_{\mu\nu}=0\eqno{(56)}$$ where, $$H_{\mu\nu}\edf\chi[
 \gamma_{\alpha\beta\mu}\gamma^{\alpha\beta}_{. .
 \nu}+\gamma_{\alpha\beta\mu}\gamma_{\nu}^{.
 \alpha\beta}+\gamma_{\alpha\beta\nu}\gamma_{\mu}^{ .
 \alpha\beta}+g_{\mu\nu}(\gamma_{\alpha\beta\sigma}\gamma^{\sigma\alpha\beta}
 -\frac{1}{2}\gamma_{\alpha\beta\sigma}\gamma^{\alpha\beta\sigma})]\eqno{(57)}$$
 and $$ f_{\mu\nu}\edf\chi[\phi_{\mu , \nu}-\phi_{\nu ,
 \mu}-\phi_{\alpha}\Lambda^{\alpha}_{ . \mu\nu}+\gamma^{. .
 \alpha}_{\mu\nu . ; \alpha}]\eqno{(58)}$$ Saez and De Juan
 [23] applied M\o ller's field equations to construct world models. They
 have obtained a number of Big Bang models depending on the choice
 of the material distribution and the value of $k$.

 Saez [24] proposed two generalization for the MTT by introducing
 a scalar field in addition of the tetrad vector field. His
 theories compose a class known as {\it "Scalar-Tetradic
 Theories of Gravity"}. He examined [25], among other things, the
 cosmological consequences of the suggested class and found that his
 result could be reduced to the corresponding results of MTT. It
 is to be noted that all world models obtained from MTT have the same
 problems as those obtained from GR.

 \subsection{Cosmology of the New General Relativity (NGR)}
 Hayashi and Shirafuji [26] constructed a theory which they called
 {\it "New General Relativity"}. The theory is constructed in the
 AP-geometry and the field equations have been derived using an
 action principle. The Lagrangian function used is in the form: $$
 \pounds\edf\lambda(\frac{R}{2k}+d_{1}(t^{\lambda\mu\nu}t_{\lambda\mu\nu})
 +d_{2}c^{\mu}c_{\mu}+d_{3}a^{\mu}a_{\nu})\eqno{(59)}$$ where
 $d_{1}$, $d_{2}$, $d_{3}$ are three parameters and $\lambda$ is
 the determinant of $\h{i}_{\mu}$. The tensor $t_{\lambda\mu\nu}$
 is defined by: $$
 t_{\lambda\mu\nu}\edf\frac{1}{2}(\Lambda_{\lambda\mu\nu}+\Lambda_{\mu\lambda\nu}-\frac{1}{6}
 (g_{\nu\lambda}c_{\mu}+g_{\mu\nu}c_{\lambda})+\frac{1}{3}g_{\lambda\mu}c_{\nu},\eqno{(60)}$$
 and the axial vector $a_{\mu}$ is defined by,
 $$a_{\mu}\edf\frac{1}{6} \epsilon_{\mu\nu\rho\sigma}
 \Lambda^{\nu\rho\sigma},\eqno{(61)}$$ and, $$
 \epsilon_{\mu\nu\rho\sigma}= \lambda
 e_{\mu\nu\rho\sigma},\eqno{(62)}$$ $e_{\mu\nu\rho\sigma}$ is the
 Levi-Civita totally anti-symmetric tensor.

 The field equations of NGR can be written in the form: $$
 G^{\mu\nu}+S^{\mu\nu}=\kappa{\tilde{T}}^{\mu\nu}\eqno{(63)}$$
 where,$$S^{\mu\nu}\edf 2kF^{{\stackrel{\mu}{}}{\stackrel{\nu}{}}{\stackrel{\lambda}{}}}
 _{.\ .\ .\ |{\lambda}}-2kc_{\mu}F^{\mu\nu\lambda}+2kK^{\mu\nu}-kg^{\mu\nu}L$$ and,
$$F^{\mu\nu\lambda}\edf
d_{1}(t^{\mu\nu\lambda}-t^{\mu\lambda\nu})-d_{2}(g^{\mu\nu}c^{\lambda}-g^{\mu\lambda}c^{\nu})
-\frac{1}{3} d_{3} \epsilon^{\mu\nu\lambda\sigma}
a_{\sigma},\eqno{(64)}$$
$$K^{\mu\nu}\edf\Lambda^{\rho\sigma\mu}F^{.\ .\nu}_{\rho\sigma}-
\frac{1}{2} \Lambda^{\nu\rho\sigma}F^{\mu}_{ . \rho\sigma},
\eqno{(65)}$$ and $$L\edf
d_{1}(t^{\lambda\mu\nu}t_{\lambda\mu\nu})+d_{2}c^{\mu}c_{\mu}+d_{3}a^{\mu}a_{\mu}.\eqno{(66)}$$
It is clear that $\tilde{T}^{\mu\nu}$ in (63) is non-symmetric
phenomenological material-energy tensor. Taking the vectorial
divergence of both sides of (63), Hayashi and Shirafuji found
that, $$ \tilde{T}^{\mu\nu}_{.\ .\ ; \nu}- \gamma^{\nu\lambda\mu}
\tilde{T}_{\nu\lambda}=0.\eqno{(67)}$$ They considered (67) as a
generalization of the law of conservation. This will reduce to
conservation in orthodox GR when $\tilde{T}^{\mu\nu}$ is
symmetric.

Mikhail et al.[27] applied the NGR field equations (63) together
with conservation (67) using the two AP-structures (48), (49).
They have assumed that the material distribution is described by a
material energy tensor of a perfect fluid, usually used in GR.
They have obtained two families of World models corresponding to
equations of state for dust and radiation. The models obtained
have the same problems of standard Big Bang cosmology, expect that
the existence of  horizons is conditional in some models.

\subsection{A Pure Geometric Approach to the Big-Bang}
Mikhail and Wanas [28] have constructed a field theory in an
attempt to unifying gravity and electromagnetism, in the context
of AP-geometry. The theory is a pure geometric one and it is
called the {\it "Generalized Field Theory "} GFT. The field
equations of this theory were obtained using variational calculus
but without using an action principle. However, the same field
equations could be obtained using an action principle [29]. In
both cases the Lagrangian used can be written in the form: $$
\pounds=\lambda(\Lambda_{\mu\alpha\nu}
\Lambda^{\alpha\mu\nu}-c_{\mu}c^{\mu}).\eqno{(68)}$$ This
Lagrangian has been constructed using certain assumptions
generalizing the Scheme of GR [15]. The field equations obtained
from this Lagrangian can be written in the form, $$E_{\mu\nu}=0,
\eqno{(69)}$$ where $E_{\mu\nu}$ is a second order non-symmetric
tensor defined in the AP-space. The symmetric part of this tensor
gives rise to the equations $$ R_{\mu\nu}-\frac{1}{2}
g_{\mu\nu}R=B_{\mu\nu},\eqno{(70)}$$ where $B_{\mu\nu}$ is a
geometric material-energy tensor defined by,
$$B_{\mu\nu}\edf\frac{1}{2}
g_{\mu\nu}(\sigma-\varpi)+\varpi_{\mu\nu}-\sigma_{\mu\nu},\eqno{(71)}$$
where the tensors on the R.H.S. of this definition are given in
Table 1. The skew- symmetric part of (69) gives rise to the
equations, $$F_{\mu\nu}=c_{\mu , \nu}-c_{\nu , \mu},\eqno{(72)}$$
where $F_{\mu\nu}$ is a second order skew symmetric tensor defined
by,
$$F_{\mu\nu}\edf\zeta_{\mu\nu}-\xi_{\mu\nu}+\eta_{\mu\nu},\eqno{(73)}$$
also, the tensor on the R.H.S. of this expression are defined in
Table 1. It is to be considered that $B_{\mu\nu}$ is subject to a
conservation condition as a consequence of (70).

Since the GFT
 theory is a pure geometric theory, a certain scheme, known as
 {\it "Type Analysis"} has been suggested [12], [30] to a
 attribute some physical meaning to the geometric objects of the
 AP-space. This scheme enables one to know, off hand, the
 capabilities of any AP-structure to represent physical systems.
 Applying this scheme to the two structures given by (48), (49),
 respectively, it is shown [31] that the first AP-structure (48)
 can represent a gravitational field within a material
 distribution, while the structure (49) is capable of representing
 a gravitational field in free space. Thus the structure (48) is
 to be used in order to construct  non-empty world model.

 The author [32] has applied the GFT field equations (69) to the
 AP-structure (48), and got a unique world model. This model is
 non-empty and has no particle horizons. It fixes a value for
 $k(=-1)$ i.e. it has no flatness problem, but it still has a
 singularity at t=0. A further advantage of using pure geometric
 theories [33] is that we do not need to impose any condition from
 outside the geometry used (e.g. equation of state) in order to
 solve the field equations.

 \subsection{Spin-Gravity Interaction and the Cosmological Parameters}
 Recently, a type of interaction between the quantum spin of a
 moving particle and the background gravitational field, is
 suggested [18]. The equation of motion of a spinning particle in
 a gravitational field is that given by (40). The parameter $b$ is
 given by, $$ b=\frac{n}{2}\alpha\gamma,\eqno{(74)}$$ where $n$ is
 a natural number, $\alpha$ is the fine structure constant and
 $\gamma$ is a dimensionless parameter of order unity. The use of
 this equation in the weak field limit, removed the discrepancy
 from the results of the COW-experiment [34]. Also, it helped in
 constructing a temporal model for SN1987A [35], which is in good
 agreement with supernovae mechanism.

 The author [36] studied the effect of the new suggested
 spin-gravity interaction on the cosmological parameters. The
 results obtained are tabulated in Table 3. Equation (40)
 indicates
 that trajectories of massless spinning particles in gravitational
 fields is spin dependent. The natural number $n$ takes the values
 0,1,2,3,....for particle with spin
 $0,\frac{1}{2},1,\frac{3}{2},...$ respectively. The null geodesic
 equation, which is a special case of (40) corresponding to $n=0$,
 is usually used in the context of GR to represent the trajectory
 of massless spinning particles in gravitational field. This means
 that, in using null geodesic, we neglect the effect of the
 spin-gravity interaction. Table 3 summarizes the results of
 calculations of the effect of this interaction, on the
 cosmological parameters. The second column gives the conventional
 values of these parameters, i.e. those values obtained when
 neglecting the interaction. The third , fourth and fifth columns
 give the effect of this interaction on values of the parameters.
\begin{center}
 {Table 3: Spin-Dependence of Cosmological Parameters}     \\
\vspace{0.5cm}
\begin{tabular}{|c|c|c|c|c|} \hline
& & & & \\ Parameter & Spin-0  & Spin-$\frac{1}{2}$ (neutrino)&
Spin-1 (photon) & Spin-2 (graviton)
\\ & & & &
\\ \hline & & & & \\ Hubble     & $H_{o}$ & $(1- \frac{\alpha}{2}) H_o $ & $(1-\alpha)H_{o}$
&  $(1- 2 \alpha)H_{o}$ \\ & & & & \\ \hline & & & & \\
 Age &  $\tau_{o}$ & $\frac{\tau_o}{(1-\frac{\alpha}{2})}$ &
$\frac{\tau_{o}}{(1-\alpha)}$ &  $\frac{\tau_{o}}{(1- 2 \alpha)}$
\\
 & & & & \\ \hline & & & & \\ Acceleration & $A_o$
 & $(1-\frac{\alpha}{2})(A_{o}-\frac{\alpha}{2}H_{o})$
 & $(1-{\alpha})(A_{o}-{\alpha}H_{o})$
 & $(1-2{\alpha})(A_{o}-2{\alpha}H_{o})$
\\ & & & & \\ \hline & & & & \\ Deceleration  &$ q_{o}$
&$\frac{(q_{o}-\frac{\alpha}{2H_{o}})}{(1-\frac{\alpha}{2})}$ &
$\frac{(q_{o}-\frac{\alpha}{H_{o}})}{(1-\alpha)}$ &
$\frac{(q_{o}-\frac{2 \alpha}{H_{o}})}{(1- 2\alpha)}$
\\ & & & & \\ \hline  & & & & \\ Matter & $\Omega_{o}$
& $\Omega_{o}$ & $\Omega_{o}$ & $\Omega_{o}$
\\ & & & &\\ \hline
\end{tabular}
\end{center}
It is clear from this Table that, the value of the matter
parameter is not affected by this interaction. This is because
both the mean cosmic density and the critical density have the
same dependence on Hubble's parameter. It is of interest to note
that, if we measure these parameters using different carriers of
cosmological information (e.g. photons and neutrino), one would
obtain a further confirmation, on the cosmological scale, of the
existence of spin-gravity interaction.

\section{TOPOLOGY AND COSMOLOGY}
All what is given, so far, in the previous sections can be
classified under the title {\it "Geometry and Cosmology"}. In
recent years, some articles appeared in periodicals connecting
topology to cosmology. Of course, one cannot give a complete
review about this topic in such a limited number of pages. But I
will focus on a single trend in this class, that is the recent
work of El Nashie, which I consider as related to the subject of
the present review. El Nashie suggested a special scheme, which
has been published in a number of papers, to understand nature.
This scheme depends mainly on a type of topology {\it "The Wild
Topology"} and a type of geometry {\it"The Noncommutative
Geometry"},  both related to 4-dimensional fusion algebra and
M.Fredmann 4- dimensional Topological spaces. Using this scheme he
was able to obtain physical results in excellent agreement with
all microphysical experiments and some macrophysical observations.
For example, he obtained the mass spectrum of quarks [37]; and an
acceptable value for the cosmic microwave back ground radiation
(CMBR) temperature [38]. Moreover, on the same bases, he obtained
convincing results concerning unification of fundamental
interactions [39], a general theory for quantum gravity [40], the
dimensions of heterotic string theory [41], and a value for the
super-symmetric quantum gravity coupling constant [42]. His main
calculations depend on the golden mean $\phi ( = \frac{\sqrt{5}-1}
{2})$.

El-Naschie's results are not only more than promising but also,
they stimulate many questions and various comments. For instance
we could ask the following questions:

(1) First one could ask a formal question about the title of this
theory. Why it is $\varepsilon ^{(\infty)}$ theory ? Why it is not
called the golden mean field theory.

(2) A More important question is that: Are we really able to
interpret any phenomena in the Universe without using evolutionary
scenarios ?

El-Naschie results, in the context of his constructed wild
topology, depend on the quantity $\phi$. And since $\phi$ is
constant, then every subsequent result will be constant ! What
does this mean ? As clear from the present review, there were two
rival scenarios used to interpret the general features of our
universe : The Big-Bang scenario, and the Steady State one. In the
first, the Universe as well as its constituents evolve. In the
Steady State scenario the constituents of the Universe are
evolving while the global characteristics of the Universe remain
the same. The observation of the CMBR-temperature, in the mid
sixties of the 20th century, ruled out the Steady State theory,
since there is no place for the CMBR in this theory, as CMBR is a
result of evolution of the universe. The most astonishing thing is
that the CMBR-temperature is obtained from El-Naschie scheme,
while everything is constant in his calculations!

Is El-Naschie working in a Steady State background and
consequently giving a new chance to this theory to revive? Let us
try to give an answer to this question. If we accept Mach's
principle, then any property of the constituents of the Universe
is a reflection of the large scale material distribution in the
Universe. And since, in the context of a Steady State model, this
distribution does not change(the material-energy density is
constant), then we get constant properties of the constituents
including masses of elementary particles and even the
CMBR-temperature. Other questions may be raised if we accept this
interpretation.

(3) Is El-Naschie theory dealing with stable configurations (in
the Universe) only? An answer to this question may throw some
light on his way of understanding nature. It is widely accepted
that stable systems were not born in this situation (stability
situation). It is usually assumed that such systems are born as
unstable systems, and gradually arrive to stability, satisfying
certain stability conditions. In this case, it seems that there is
no escape from evolution, the assumption which does not exist in
El-Naschie treatment.

(4) Is it possible to construct a general (or let us say,
universal) stability theory, which can transfer unstable
configurations to El-Naschie stable systems ? If yes, the
stability conditions of this theory would be algebraic (equations
or inequalities) rather than differential, in order to be
consistent with the $\varepsilon ^{(\infty )}$-theory. If this is
done, it would be considered as a complement of El-Naschie theory.
Moreover, the golden mean $\phi $ would be a real root(or the real
root)necessary to satisfy such conditions. This would construct an
acceptable bridge between existing physics and El-Naschie {\it "
Topologization of Physics"} .

\section{DISCUSSION AND CONCLUDING REMARKS}
This brief review gives alternative treatments of theoretical
cosmology. In particular it gives alternative theoretical
treatments leading to Big-Bang or Steady State cosmologies. The
standard treatment in theoretical cosmology can be summarized in
the following steps:

(1) Riemannian geometry is assumed to give a complete
representation of the physical World, including space-time.

(2) Certain Riemannian structures (e.g. (2)), satisfying some
conditions (e.g. the cosmological principle) are to be used as
basic structures for constructing World models.

(3) The equations of GR, written in Riemannian geometry, represent
good constraint  connecting the material distribution in the
Universe and the geometric structure used to describe it.

Problems of standard cosmology motivated investigators to change
the conventional treatment, summarized above, in the hope that
this change may remove one or more of these problems. The use of a
more wider geometry, the AP-geometry, in place of GR represents a
change in the first step (this is done in Section 2). Consequently
the use of the structures (48) and (49) in place of (2) represents
a change in the second step (which is given in Section 3). Finally
the use of the equations of alternative field theories, different
from GR, is a change of the third step (as presented in Section
4).

In the following we are going to compare the results obtained,
from the alternative treatments given in the present work with
standard theoretical cosmology. (note: one can reconstruct GR in
AP-geometry by taking Ricci tensor as defined in Table 1, and the
geodesic (or null geodesic) is obtained from the AP-path (40) upon
taking $n=0$)

(1) In order to compare the Big Bang results of alternative field
theories, written in the AP-geometry, with those of GR, we first
give a brief comparison of these theories in Table 4.
\begin{center}
Table 4: Comparison Between Field Theories Giving Big Bang Models
\vspace{1.3cm}
\begin{tabular}{|c|c|c|c|c|c|} \hline
& & & & & \\ Field & Reference & Field & Field & Gravitational &
$T_{\mu\nu}$ \\ Theory & &  Equations & Variables & Potential & \\
& & & & & \\ \hline & & & & & \\ GR (1916) & cf.[1] &
$G_{\mu\nu}=-\kappa T_{\mu\nu}$ & $g_{\mu\nu}$ & $g_{\mu\nu}$ &
Phenom. \\ & & & & & \\  \hline & & & & & \\ GFT (1977) & [28] &
${G}_{\mu\nu} = B_{\mu\nu}$ & $\h{i}_\mu$ & $g_{\mu\nu}$ & Geomet.
\\ & & $F_{\mu\nu}=c_{\mu , \nu}-c_{\nu , \mu}$ & & & \\ & & & & &
\\ \hline & & & & & \\ MTT (1978) & [23] &
$G_{\mu\nu}+H_{\mu\nu}=-\kappa T_{\mu\nu}$ & $\h{i}_\mu$ &
$g_{\mu\nu}$ & Phenom. \\ & & $f_{ \mu\nu }=0$ & & & \\ & & & & &
\\ \hline & & & & & \\ NGR (1979) & [26] &
$G_{\mu\nu}+S_{\mu\nu}=-\kappa \tilde{T}_{\mu\nu}$ & $\h{i}_\mu$ &
$g_{\mu\nu}$ & Phenom. \\ & & & & & \\ \hline
\end{tabular}
\end{center}
The last column of this table indicates whether the material-
energy tensor is phenomenological or geometric.

Table 5 gives a comparison between the Big Bang cosmology
resulting from the alternative theories given in the present work
and that resulting from GR. It is of interest to note that the
exisdence of horizon in the NGR is conditional. It is clear from
this Table that in case of GFT, the horizon and flatness problems
disappeared from the model obtained. Also this model is a unique
one and has the advantage that extra conditions (equation of
state) is not needed to construct the model, but a relation
between the pressure and density is obtained as a result of this
model. This is the main advantage of using a pure geometric
approach to cosmology [33].
\newpage

\begin{center}
Table 5: Comparison between the resulting Big Bang cosmologies
\vspace{.3cm}
\begin{tabular}{|c|c|c|c|c|} \hline
 & & & & \\
Criterion & GR, cf[1] & GFT, [32]& MTT, [22]& NGR, [27]   \\
\hline
 & & & & \\
 Space    & Riemannian &  AP-space & AP-space & AP-space \\
&   & & &  \\ \hline Free &  & & &    \\ parameters &   No & No &
One  & Three  \\
 & & &  & \\
\hline
 & & & & \\
Energy-  &     Symmetric &  Symmetric &  Symmetric & Non-symmetric
\\ momentum &  Phenomological  & Geometric & Phenomological  &
Phenomological \\ tensor $T^{\mu \nu}$  & tensor &   tensor &
tensor         &      tensor \\ & & & &  \\ \hline
 & & & &  \\
Basic  &  Homogenity &  Homogenity &  Homogenity & Homogenity \\
assuption & {\&} Isotropy & {\&} Isotropy & {\&} Isotropy & {\&}
Isotropy \\ & &  & &  \\ \hline
 & & & & \\
k for non-static,&  $+1,0,-1$ & $-1$ & $+1,0,-1$ & $+1,0,-1$ \\
non-empty models &   & & &               \\ & & & &  \\ \hline
 & & & & \\
Number of & & &  & \\ models allowed & Many & One & Many & Many \\
& &  & &  \\ \hline
 & & & & \\
Particle & & & & \\ Horizons & Yes & No & Yes & Conditional \\ & &
& &  \\ \hline
 & & & & \\
Flatness & & & & \\ Problem & Yes & No & Yes & Yes \\ & & & &  \\
\hline & & & & \\ Need for extra condition &equ. of state &- &
Equ.of state & Equ.of state \\ & & & & \\ \hline
\end{tabular}
\end{center}
\newpage
(2) Also to compare Steady State results of alternative theories
constructed in AP-geometry with those resulting from the modified
equations of GR(6), we establish Table 6.
\begin{center}
Table 6: Comparison between the resulting Steady State cosmologies
\vspace{.3cm}
\begin{tabular}{|c|c|c|c|} \hline
 & & &  \\
Criterion & Hoyle [6] & McCrea and Mikhail [19]& Mikhail [21]
\\ \hline
 & & &  \\
 Space    & Riemannian &  AP-space & AP-space  \\
 \hline
 &  & &  \\ Creation tensor &   Phenom. & Geometric &?
\\ \hline
 & & &  \\
Extra conditios  &   Equation of state & Equation of state & -
  \\ \hline
 & & &   \\
$k$  & 0 &  0 & +1, 0, -1 \\ \hline
\end{tabular}
\end{center}
It is to be considered that all the Steady State results listed in
Table 6 are free from singularity and horizon problem. Further
more the use of a pure geometric approach ( fourth column of the
table) indicates that their is no need for an equation of state to
construct a world model.

(3) Some authors [23] construct World models without using the
AP-structures (48),(49). However it is more appropriate to use
(48) or(49) in order to guarantee homogeneity and isotropy in the
general case. The use of the roots of the metric tensor to produce
tetrads is not sufficient to obtain the most general AP-structure
satisfying the cosmological principle. This is because the ten
components of the metric tensor $g_{\mu\nu}$ can not fix the
sixteen tetrad components $\h{i}_\mu$ uniquely.

(4) As in subsection 2.2, the PAP-geometry is more wider than the
conventional AP and the Riemannian geometries. It has sufficient
structure for other physical interactions to be represented,
beside gravity. For example its general path equation (40) are
used [18] to describe the trajectories of massless spinning
particles in a background gravitational field. Its R.H.S. is
interpreted as representing a type of interaction between the
quantum spin of the moving particle and the gravitational field.
Since particle carrying the cosmological information are massless
spinning particles, then their trajectories in the cosmic space
will be affected by the spin-gravity interaction. This will affect
information carried by such particles, and then it will be of
interest to know how to free information from this interaction.
This is clear from Table 3 which gives the effect of this
interaction on the cosmological parameter. This is one of the
achievements of using the non-conventional approach given in the
present review.

(5) El-Nashie [43] quoted two arguments to support, what I am calling, topologization of physics. The first is the relation between the
golden mean and the Fibonacci series. The second is the relation between the Hausdorff dimensions of the Menger sponge
and the CMBR-temperature. Let us discuss the consequences of these
two arguments.

The Fibonacci series (cf. [44]) (0, 1, 1, 2, 3, 5, 8, ...) is a
series in which each term $ t_{n+1}$ is the sum of the preceding
two terms $(t_{n} + t_{n-1})$. The golden mean is obtained as: $$
\phi = \lim \limits_{n\to \infty } \frac{t_{n}}{t_{n+1}} $$ By this definition, although
$\phi$ is not an exact number, it converges to a constant.
Consequently, everything in El-Nashie theory would be constant, or
converges to a constant. There is no direct time evolution as stated in
Section 5, but there is another type of evolution i.e. the Fibonacci gross law.

The second argument gives a further confirmation to the above
remark. The CMBR -temperature has a strong relation, via entropy
and complexity theory, with the Hausdorff dimensions of the Menger
sponge, which is given by $ \frac {log 20} {log 3} = 2.7268...$.
This dimension, although it is not exact, it converges to a
constant value without any time evolution. So, if there is any
causal relation between the CMBR- temperature and the Hausdorff
dimensions of the Menger sponge, then one of El-Nashie important
results [38] is a consequence of using a constant! Is there any
physical relationship between this argument and the Zeldovich idea
that our Universe is similar to a sponge? El Nashie states that
there is [43].

The conclusion is that El-Nashie theory may need a general
stability theory (may be a generalization of a theory of the type
of the KAM theory) in order to take over unstable systems to El-
Nashie stable systems. In this case, once again, we believe that El-Nashie is
working in a Steady State background.

\section*{References}

{[1] Weinberg, S. (1972) {\it "Gravition and Cosmology "}}, John Wiley \& Sons.  \\
{[2] Robertson, H. P. (1935) Ap. J., {\bf 82}, 284.

~~~~~~~~~~~~~~~~~~~~~~~(1936) Ap. J., {\bf 83}, 187, 257.}
\\ {[3] Walker, A. G. (1936) Proc. Lond. Math. Soc., {\bf 42}, 90. } \\ {[4] Bondi, H. and Gold, T.
(1948) Mon. Not. Roy. Astr. Soc. , {\bf 108} , 252.} \\ {[5]
Hoyle, F. (1948) Mon. Not. Roy. Astr. Soc, {\bf 108}, 372.} \\
{[6] Hoyle, F. (1949), Mon. Not. Roy. Astr. Soc. {\bf 109}, 365.}
\\ {[7] Hoyle, F. and Narlikar, J. (1963) Proc. Roy. Soc., {\bf
A273},1 .}
\\ {[8] Hoyle, F. and Narlikar, J. (1966) Proc. Roy. Soc. {\bf
A290}, 177.}
\\ {[9] Levi-Civita, T. (1952), {\it A simlified Presentation of
Einstein's Unified Field Equations},

A translated pamphlet, Blackie and Son Limited }.
\\ {[10] Mikhail, F. I. (1952) Ph. D. Thesis, London University.}
\\ {[11] Wanas, M. I. (2000) Turk. J. Phys. {\bf 24}, 473.} \\
{[12] Wanas, M. I. (2000) Proc. Romanian Conf. on{\it "Finsler,
Lagrange and Hamilton}

{\it Geometries" } P. 284, eds V. Blanvta and Gh Neaga.}
\\{[13]
Robertson, H. P. (1932) Ann. Math. Princeton, {\bf 33}, 496.}\\
{[14] Mikhail, F. I. (1962), Ain Shams Sci. Bull. {\bf 6}, 87.}
\\ {[15] Wanas, M. I. (1975) Ph. D. Thesis, Cairo University,
Egypt.}
\\ {[16] Wanas, M. I., Melek, M. and Kahil, M. E. (1995)
Astrophy. Space Sci.{\bf 228}, 273.}
\\ {[17] Wanas, M. I. and Melek, M. (1995) Astrophysics Space
Sci., {\bf 228}, 277.} \\ {[18] Wanas, M. I. (1998) Astrophysics
Space Sci.{\bf 258}, 237.}  \\ {[19] McCrea, W.H. and Mikhail, F.
I. (1956) Proc. Roy. Soc. Lond {\bf A235}, 11.}
\\ {[20] Mikhail, F. i. (1964) Il Navo Cimento, X {\bf 32}, 886.}
\\ {[21] Mikhail, F. I. (1967) Ain Shams Sci. Bull. {\bf 11}, 301.}
\\ {[22] M\o ller, C. (1978) Math. Fys. Medd. Dan. Vid. Selek. {\bf
39}, 1}. \\ {[23] Saez, D. and de Juan, T. (1984) Gen. Rel. Grav.
{\bf 16}, 501.} \\ {[24] Saez, D. (1983) Phys. Rev. {\bf D27},
2839.}
\\ {[25] Saez, D. (1985) Gen. Rel. Grav. {\bf 17}, 579.} \\ {[26]
Hayashi, K. and Shirafuji, T. (1979) Phys. Rev. {\bf D19}, 12.}
\\ {[27] Mikhail, F. I., Wanas, M. I. and Nashed, G. G. L. (1995)
Astrophys. Space Sci

{\bf 228}, 225.} \\ {[28] Mikhail, F.I. and
Wanas, M.I. (1977) Proc. Roy. Soc. Lond. {\bf A356} , 471.} \\
{[29] Mikhail, F. I. and Wanas, M. I. (1998) gr-qc/{\bf 9812086}.}
\\
 {[30] Mikhail, F.I. and Wanas, M.I. (1981) Int. J.
Theoret. Phys., {\bf 20}, 671.} \\ {[31] Wanas, M. I. (1986)
Astrophys. Space. Sci., {\bf 127}, 21.} \\ {[32] Wanas, M. I.
(1989) Astro. Space Sci., {\bf 154}, 165}. \\{[33] Wanas, M. I.
(1999) Proc.IAU-Symp \# 183, 316, ed. K.Sato, Kluwer Pub. } \\{[34]
Wanas, M.I., Melek, M. and Kahil, M.E. (2000), Gravit.Cosmol. {\bf
6}, 319.} \\
 {[35] Wanas, M.
I. , Melek, M. and Kahil, M. E. (2001) Proc.IAU-Symp \# 205. }\\
{[36] Wanas, M. I. (2001) Proc.IAU-Symp \# 201. }
\\ {[37] El-Naschie, M. S. (2002)
chaos, solutions and fractals {\bf \ 14}, 369.}
\\ {\ [38] El-Naschie, M. S. (2002) invited talk, coll.\#2 on
{\it "Astronomy and space Researches"},

held in Cairo, Egypt, April 2002.} \\ {\ [39] El-Naschie, M. S.
(2000) chaos, solutions and fractals, {\bf 11}, 1149.} \\
{\ [40] El-Naschie, M. S. (2001) chaos, solutions and fractals, {\bf 12}, 875.%
} \\ {\ [41] El-Naschie, M. S. (2001) chaos, solutions and
fractals, {\bf 12}, 1025.} \\ {\ [42] El-Naschie, M. S. (2001)
chaos, solutions and fractals, {\bf 12}, 1361.}\\
 {\ [43]
El-Naschie, M. S. (2001)Private communication, see also his
article in this volume.}
\\ {\ [44] Stewart, I. (1997) {\it Does God Play Dice?} 2nd. ed.,
Penguin.}

\end{document}